\documentclass[12pt]{article}
\usepackage{graphicx}
\usepackage{dcolumn}
\usepackage{amsmath}
\usepackage{units}
\usepackage{pstricks}

\newcommand{\BABARPubYear}    {04}

\newcommand{\BABARConfNumber} {30}
\newcommand{\SLACPubNumber} {10617}

\input{pubboard/babarsym}

\newcommand{\Btag}{\ensuremath{\B_\mathrm{tag}}}
\newcommand{\Bztojpsiks}{\ensuremath{\Bz\to\jpsi\KS}}
\newcommand{\Bztophiks} {\ensuremath{\Bz\to\phi\KS}}
\newcommand{\Bztoetapks} {\ensuremath{\Bz\to\eta'\KS}}

\newcommand{\Bztokpkmks} {\ensuremath{\Bz\to K^+ K^- \KS}}
\newcommand{\Bztokspiz} {\ensuremath{\Bz \to \KS\piz}}
\newcommand{\Bztokzpiz} {\ensuremath{\Bz \to K^{0}\piz}}
\newcommand{\ckspiz} {\ensuremath{C_{\KS\piz}}}
\newcommand{\skspiz} {\ensuremath{S_{\KS\piz}}}
\def\cf {\ensuremath{C_f}}
\def\sf {\ensuremath{S_f}}

\newcommand{\Brec}{\ensuremath{B_{\CP}}}
\newcommand{\Ups}{\ensuremath{\Upsilon}}
\newcommand{\dte}{\ensuremath{\sigma(\deltat)}}
\newcommand{\thetacms}{\ensuremath{\theta_{B}^*}}
\newcommand{\costhetacms}{\ensuremath{\cos\thetacms}}
\newcommand{\mmiss}{\ensuremath{m_\text{miss}}}
\newcommand{\mb}{\ensuremath{m_{B}}}
\newcommand{\rhom}{\ensuremath{\rho^{-}}}
\newcommand{\sPlot}{s-Plot}

\newcommand{\babarprelim}{
  \begin{picture}(0,0)(0,0)
    \rput[ct](0,0){\bf\babar}
    \rput[ct](0,-0.3) {\bf\tiny preliminary}
  \end{picture}
  }

\long\def\inst#1{\par\nobreak\kern 4pt\nobreak
    {\it #1}\par\vskip 10pt plus 3pt minus 3pt}

\begin{document}
{\pagestyle{empty}

\begin{flushright}
  \babar-CONF-\BABARPubYear/\BABARConfNumber \\
  SLAC-PUB-\SLACPubNumber \\
\end{flushright}

\par\vskip 5cm

\begin{center}
  \Large \bf \boldmath Measurements of the Branching Fraction 
  and \CP-Violating Asymmetries of \Bztokspiz{} Decays
\end{center}
\bigskip

\begin{center}
\large The \babar{} Collaboration\\
\mbox{ }\\
\today
\end{center}
\bigskip \bigskip

\begin{abstract}
  We present measurements of the branching fraction and
  time-dependent \CP-violating (CPV) asymmetries in \Bztokspiz{} decays
  based on $227$ million $\Y4S\to\BB$ decays collected with the \babar{} 
  detector at the PEP-II asymmetric-energy $B$ factory at SLAC. We  
  obtain a branching fraction 
  $\BR{}(\Bztokzpiz) = (11.4 \pm 0.9 \pm 0.6)\cdot 10^{-6}$,
  the magnitude of the direct CPV asymmetry 
  $\ckspiz = 0.06 \pm 0.18 \pm 0.06$ and the magnitude of the CPV asymmetry in the
  interference between mixing and decay $\skspiz = 0.35^{+0.30}_{-0.33} \pm 0.04$, 
  where the first error is statistical and the second systematic.
  All results are preliminary.
\end{abstract}

\date{\today}

\vfill
\begin{center}
  Submitted to the 32$^{\rm nd}$ International Conference on High-Energy Physics, ICHEP 04,\\
  16 August---22 August 2004, Beijing, China
\end{center}

\vspace{1.0cm}
\begin{center}
{\em Stanford Linear Accelerator Center, Stanford University, 
Stanford, CA 94309} \\ \vspace{0.1cm}\hrule\vspace{0.1cm}
Work supported in part by Department of Energy contract DE-AC03-76SF00515.
\end{center}

\newpage
}

\begin{center}
\small

The \babar\ Collaboration,
\bigskip

%
B.~Aubert,
R.~Barate,
D.~Boutigny,
F.~Couderc,
J.-M.~Gaillard,
A.~Hicheur,
Y.~Karyotakis,
J.~P.~Lees,
V.~Tisserand,
A.~Zghiche
\inst{Laboratoire de Physique des Particules, F-74941 Annecy-le-Vieux, France }
% ...

\end{center}\newpage

\section{INTRODUCTION\label{sec:Introduction}}

In the Standard Model (SM) the time-dependent \CP{} symmetry violation
in \Bz{} decays via $\b\to\s\qbar\q$ ($\q\in\{u,d,s,c\}$) transitions is
governed by the Cabibbo-Kobayashi-Maskawa (CKM) phase
$\beta\equiv\arg(-V_{cd}V^*_{cb}/V_{td}V^*_{tb})$~\cite{CKM}.
Measurements of \stwob{} in tree-level $\b\to\s\cbar\c$
transitions have established the first experimental evidence for \CP{}
violation in the $B$ meson
system~\cite{BaBarSin2betaObs,BelleSin2betaObs}. The combined result
of the latest measurements of \stwob{} in these transitions is
in agreement with SM predictions~\cite{Eidelman:pdg2004}.

The measurement of \stwob{} in decays dominated by a penguin
loop-level $\b\to\s\qbar\q$ transition is particularly interesting
because of the sensitivity to physics beyond the Standard 
Model~\cite{ref:newphysics}.  The \B{} factory
experiments have explored time-dependent CPV asymmetries in several
such decays~\cite{ref:cc}, namely \Bztophiks{}~\cite{Abe:2003yt,Aubert:2004ii},
\Bztoetapks{}~\cite{Abe:2003yt,Aubert:2003bq},
\Bztokpkmks{}~\cite{Abe:2003yt,Aubert:2004ta}, $\Bz\to
f^{0}\KS$~\cite{Aubert:2004am} and \Bztokspiz{}~\cite{Aubert:2004xf}.

In this paper we present an update of our previous measurement of
\CP{} violation in \Bztokspiz{} decays, reported in
reference~\cite{Aubert:2004xf}. The CKM
and color suppression of the tree-level $b\to s\bar{u}u$ transition
leads to the expectation that this decay is dominated by a
top-quark-mediated $\b\to\s\dbar\d$ penguin amplitude, which carries a
weak phase $\arg(V_{\t\b}V_{\t\s}^*)$. If other contributions, such as
the $\b\to\s\u\ubar$ tree amplitude, are ignored, the time-dependent
CPV asymmetry is governed by \stwob{}~\cite{Fleischer:1995cg}. The
bound on the deviation from \stwob{} due to Standard Model
contributions with a different weak phase is $\sim0.2$ from SU(3)
flavor symmetry~\cite{Gronau:2003kx} and $\sim0.1$ in model-dependent QCD
calculations~\cite{ref-qcdbounds}.

In addition to the CPV parameters we also present an update of our
previous measurement of the branching fraction of
\Bztokspiz{}~\cite{Aubert:2003sg}.  Existing experimental data on
branching fractions for $B \to K \pi$ decays show a small discrepancy
with respect to various calculations in the literature, the so-called
`$K\pi$ puzzle'~\cite{ref-kpipuzzle1,ref-qcdbounds}.
In particular, the ratio of the \Bztokspiz{} branching fraction to the
other $B \to K \pi$ branching fractions is about $2$ standard
deviations larger than inferred from isospin symmetry. Further
experimental input might either resolve the puzzle or provide evidence
for new physics~\cite{Buras:2003dj}.

\section{THE \babar\ DETECTOR AND DATASET}

The results are based on a sample of $226.6\pm2.5$
million $\Y4S\to\BB$ decays collected in 1999-2004 with the \babar{}
detector at the PEP-II $\epem$ energy-asymmetric collider at
the Stanford Linear Accelerator Center.  The \babar{} detector, 
fully described in~\cite{ref:babar}, provides charged particle
tracking through a combination of a five-layer double-sided silicon
micro-strip detector (SVT) and a 40-layer central drift chamber (DCH),
both operating in a \unit[1.5]{T} magnetic field. Charged kaon and
pion identification is
achieved through measurements of particle energy-loss ($dE/dx$) in the
tracking system and Cherenkov angle ($\theta_c$) in a detector of
internally reflected Cherenkov light (DIRC).  A segmented CsI(Tl)
electromagnetic calorimeter (EMC) provides photon detection and
electron identification.  Finally, the instrumented flux return (IFR)
of the magnet allows discrimination of muons from pions.

\section{ANALYSIS METHOD}

At the \Y4S{} resonance, time-dependent CPV asymmetries are measured
by reconstructing the distribution of the difference of the proper
decay times, $\deltat\equiv t_{\CP}-t_\text{tag}$, where the $t_{\CP}$
refers to the decay time of the signal \Bz{} and $t_\text{tag}$ to the
other \B{} ( \Btag). If \Bz{} and \Bzb{} decay to a common \CP{}
eigenstate $f$, the \deltat{} distribution follows
\begin{equation}
  \label{eqn:td} 
  {\cal P}^{\Bz}_{\Bzb}(\deltat) \; = \; \frac{e^{-|\deltat|/\tau}}{4\tau} \times
  \left[ \: 1 \; \pm \; 
    \left( \: S_f \sin{( \deltamd\deltat)} - C_f \cos{( \deltamd\deltat)} 
      \: \right) \: \right] \; ,
\end{equation}
where the upper (lower) sign corresponds to \Btag{} decaying as \Bz{}
(\Bzb), $\tau$ is the \Bz{} lifetime averaged over the two mass
eigenstates, \deltamd{} is the mixing frequency, $C_f$ is the
magnitude of direct \CP{} violation and $S$ the magnitude of \CP{}
violation in the interference between mixing and decay. For the case
of pure penguin dominance, we expect $S_{\KS\piz}=\stwob$, and
$C_{\KS\piz}=0$.

We search for \Bztokspiz{} decays in \BB{} candidate events 
selected using charged particle multiplicity and event
topology~\cite{ref:Sin2betaPRD}.  We reconstruct $\KS\to\pip\pim$
candidates from pairs of oppositely charged tracks.  The two-track
combinations must form a vertex with a $\chi^2$ consistency larger
than $0.001$, a $\pip\pim$ invariant mass within \unit[$11.2$]{\mevcc}
($\sim3.5\sigma$) of the nominal \KS\ mass~\cite{Eidelman:pdg2004} and
a reconstructed decay length greater than five times its uncertainty.
We form $\piz\to\gamma\gamma$ candidates with an invariant mass
\unit[$110 < m_{\gamma\gamma} < 160$]{\mevcc} from pairs of photon
candidates in the EMC that are isolated from any charged tracks, carry
a minimum energy of \unit[$50$]{\mev}, and have the expected
lateral shower shapes. \Bztokspiz{} candidates are reconstructed from \KS{}
\piz{} combinations and constrained to originate from the 
\epem{} interaction point using a geometric fit. We require that the
$\chi^2$ consistency of the fit, which has one degree of freedom, be
larger than $0.001$.
 
For each $B$ candidate we compute two independent kinematic variables,
namely the invariant mass \mb{} and the missing mass $\mmiss = |
q_{\epem} - \hat{q}_B|$, where $q_{\epem}$ is the four-momentum of the
initial \epem{} system and $\hat{q}_B$ is the four-momentum of the
\Bztokspiz{} candidate after a mass constraint on the \Bz{} is
applied. Compared to the kinematic variables $\DeltaE = E^{*}_B -
\frac{1}{2}\sqrt{s}$ and $\mes = \sqrt{\frac{1}{4} s - p^{*2}_B}$
(where $\sqrt{s}=|q_{\epem}|$ and the asterisk denotes the center of
momentum frame) that are traditionally used to select $B$ decays in
\babar{}, the present combination of variables exhibits a smaller
correlation and a better background suppression, in particular for
modes with a relatively poor $B$ energy resolution.  We select
candidates with \mb{} within \unit[$150$]{\mevcc} of the nominal \Bz{}
mass~\cite{Eidelman:pdg2004} and \unit[$5.11<\mmiss<5.31$]{\gevcc},
which includes a sideband region for background characterization.

To discriminate jet-like $\epem\to\qqbar$ events (with
$q\in\{u,d,s,c\}$) from the more uniformly-distributed \BB{} events we
exploit the ratio $L_{2}/L_{0}$ of two Legendre moments defined as
$L_j\equiv\sum_i |{\bf p}^*_i| |\cos \theta^*_i|^j$, where ${\bf
  p}^*_i$ is the momentum of particle $i$ in the \epem{} rest frame
and $\theta^*_i$ is the angle between ${\bf p}^*_i$ and the thrust
axis of the \Bz{} candidate. We require $L_2/L_0<0.55$, which
suppresses the background by more than a factor $3$ at the cost of
approximately \unit[$10$]{\%} in signal efficiency.  Finally, we
require $|\costhetacms|<0.9$, where $\thetacms$ is the angle between
the \Bz{} candidate momentum and the $e^{+}$ momentum in
the \epem{} rest frame. For \B{} candidates the distribution of
$\thetacms$ follows ${\cal P}(\costhetacms) = 1 - \cos^2\!\thetacms$,
whereas for continuum events it is nearly flat.  After all
selections the average candidate multiplicity in events
with at least one candidate is $\sim 1.007$.  We select the candidate
with the smallest $\chi^2$ on the \piz{} mass as computed 
in the \Bz{} candidate vertex fit.

For each \Bztokspiz{} candidate we examine the remaining tracks and
neutral candidates in the event to determine the decay vertex position
and the flavor of \Btag. Using a neural network based on kinematic and particle
identification information~\cite{ref:sin2betaPRL02} each event is
assigned to one of five mutually exclusive tagging categories,
designed to combine flavor tags with similar performance and \deltat{}
resolution.  We parameterize the performance of this algorithm in a
data sample ($B_{\rm flav}$) of fully reconstructed $\Bz\to
D^{(*)-}\pip/\rho^+/a_1^+$ decays. The average effective tagging
efficiency obtained from this sample is $Q = \sum_c \epsilon_S^c (1-2w^c)^2=0.288\pm 0.005$,
where $\epsilon_S^c$ and $w^c$ are the efficiencies and mistag
probabilities, respectively, for events tagged in category
$c\in\{1\cdots 5\}$. For the background the fraction of events
($\epsilon_B^c$) and the asymmetry in the rate of $\Bz$ versus $\Bzb$
tags in each tagging category are extracted from a fit to the data.

To compute the proper time difference \deltat{} the \Btag{} vertex is
inclusively reconstructed from the remaining charged particles in the
event using the trajectory derived from the reconstruction of the
\Brec{} candidate as a seed~\cite{ref:Sin2betaPRD}. The time
difference \deltat{} and its uncertainty are extracted with a global
fit to the $\Ups(4S)\to\Bz\Bzb$ decay tree that takes the information
on the beam energy and the position of the interaction point (IP) into
account.
The position and size of the interaction region are determined on a
run-by-run basis from the spatial distribution of vertices from
two-track events.  The uncertainty in the IP position follows from the
size of the interaction region (about \unit[$200$]{$\mu$m} horizontal
and \unit[$4$]{$\mu$m} vertical).

Without additional constraints the single \KS{} trajectory emerging
from the \Bztokspiz{} decay vertex provides insufficient information
on the $\Bz$ vertex position for a meaningful \deltat{} measurement.
To obtain the required resolution we constrain the sum of the two $B$
lifetimes in the $\Ups(4S)\to\Bz\Bzb$ fit ($t_{\CP}+t_\text{tag}$) to
be equal to $2\:\tau_{\Bz}$ with an uncertainty
$\sqrt{2}\:\tau_{\Bz}$.  We have verified in a Monte Carlo simulation
that this procedure leads to an unbiased estimate of \deltat{}. 

For the \unit[$\sim40$]{\%} of events in which each of the 
two pion candidates from the
\KS{} decay does not have at least 4 SVT hits or for which
\unit[$\dte>2.5$]{ps} or \unit[$\deltat>20$]{ps}, the \deltat{}
information is not used. However, since $C$ can also be extracted from
flavor tagging information alone, these events still contribute to the
measurement of $C$.

We extract the signal yield and CPV parameters from an unbinned
maximum-likelihood fit to $m_B$, \mmiss{}, $L_{2}/L_{0}$,
\costhetacms{}, \deltat{} and the flavor tag variables. Exploiting
sideband regions in data for the background and 
Monte Carlo simulation for the signal, 
we have verified that with the selection presented above
these observables are sufficiently independent that we can construct
the likelihood from the product of one dimensional probability density
functions (PDFs).  The PDFs for signal events are parameterized from
either more copious fully-reconstructed $B$ decays in data or from
simulated events.  For background PDFs we select the functional form
from data in the sideband regions of the other observables, in which 
backgrounds dominate.  We include these regions in the fitted sample
and simultaneously extract the parameters of the background PDFs along
with the CPV measurements.

We obtain the PDF for the \deltat{} of signal decays from the
convolution of Eq.~\ref{eqn:td} with a resolution function ${\cal
  R}(\delta t \equiv \deltat -\deltat_{\rm true},\sigma_{\deltat})$.
The resolution function is parameterized as the sum of a `core' and a
`tail' Gaussian, each with a width and mean proportional to the
reconstructed $\sigma_{\deltat}$, and a third Gaussian centered at
zero with a fixed width of \unit[$8$]{ps}~\cite{ref:Sin2betaPRD}.  We
have verified in simulation that the parameters of ${\cal R}(\delta t,
\sigma_{\deltat})$ for \Bztokspiz\ decays are similar to those
obtained from the $B_{\rm flav}$ sample, even though the distributions
of $\sigma_{\deltat}$ differ considerably. We therefore extract these
parameters from a fit to the $B_{\rm flav}$ sample.  We find that the
\deltat{} distribution of background candidates is well described by a
delta function convolved with a resolution function with the same
functional form as used for signal events. The parameters of the
background function are determined in the fit.

To extract the yield and the CPV asymmetries we maximize the logarithm
of the extended likelihood 
\small\begin{eqnarray*} 
  \lefteqn{{\cal L}(\sf,\cf,N_S,N_B,f_S,f_B,\vec{\alpha}) = 
    \frac{e^{-(N_S+N_B)}}{(N_S+N_B)\,!} \times } \\
  & & \prod_{i \in I} \left[ N_S f_S
      \epsilon^{c}_S{\cal P}_S(\vec{x}_i,\vec{y}_i;\sf,\cf) +
      N_B f_B \epsilon^{c}_B {\cal P}_B(\vec{x}_i,\vec{y}_i;\vec{\alpha}) \right] \times \\
 & & \prod_{i \in II} \left[ N_S (1-f_S)
    \epsilon^{c}_S {\cal P}'_S(\vec{x}_i;\cf) + N_B (1-f_B)
    \epsilon_B^{c} {\cal P}'_B(\vec{x}_i;\vec{\alpha}) \right],
\end{eqnarray*}
where $I$ ($II$) is the subset of events with (without) \deltat{} information. The
probabilities ${\cal P}_S$ and ${\cal P}_B$ are products of PDFs for
signal ($S$) and background ($B$) hypotheses evaluated for the
measurements
$\vec{x}_i=\{\mb,\mmiss,L_{12}/L_{10},\costhetacms,\text{tag},\text{tagging
  category}\}$ and $\vec{y}_i=\{\deltat,\sigma_{\deltat}\}$. Along
with the CPV asymmetries \sf{} and \cf{}, the fit extracts the yields
$N_S$ and $N_B$, the fractions of events with $\deltat$ information
$f_S$ and $f_B$, and the remaining parameters, collectively denoted by
$\vec{\alpha}$. These include all parameters of background PDFs
and some parameters of the signal PDFs, such as the mean values of
\mb{} and \mmiss{}.

\section{RESULTS\label{sec:Physics}}

Fitting the data sample of $9726$ \Bztokspiz{} candidates, we find
\mbox{$N_S=300\pm 23$} signal decays with 
\begin{equation*}
  \skspiz \; = \; 0.35 \:^{+0.30}_{-0.33} \: \text{(stat)} \: \pm \: 0.04 \: \text{(syst)}
\end{equation*}
and
\begin{equation*}
  \ckspiz \; = \; 0.06\: \pm \: 0.18 \: \text{(stat)} \: \pm 0.06 \: \text{(syst)} \;.
\end{equation*}
The total detection efficiency for \Bztokspiz{} decays with
$\KS\to\pip\pim$ and $\piz\to\gamma\gamma$ is
\unit[$0.34\pm0.02$]{\%}. With the \KS{} and \piz{} branching
fractions taken from~\cite{Eidelman:pdg2004} and assuming equal
production of charged and neutral $B$ mesons at the \Y4S{} resonance,
we obtain a branching fraction 
\begin{equation*}
  \BR{}(\Bztokzpiz) \; = \; (\,11.4 \: \pm \: 0.9 \: \text{(stat)} \: \pm \: 0.6
  \: \text{(syst)} \: ) \; \times \; 10^{-6} \; .
\end{equation*}
The evaluation of the systematic uncertainties is described below.

\begin{figure}[!tbp]
  \includegraphics[width=0.49\linewidth]{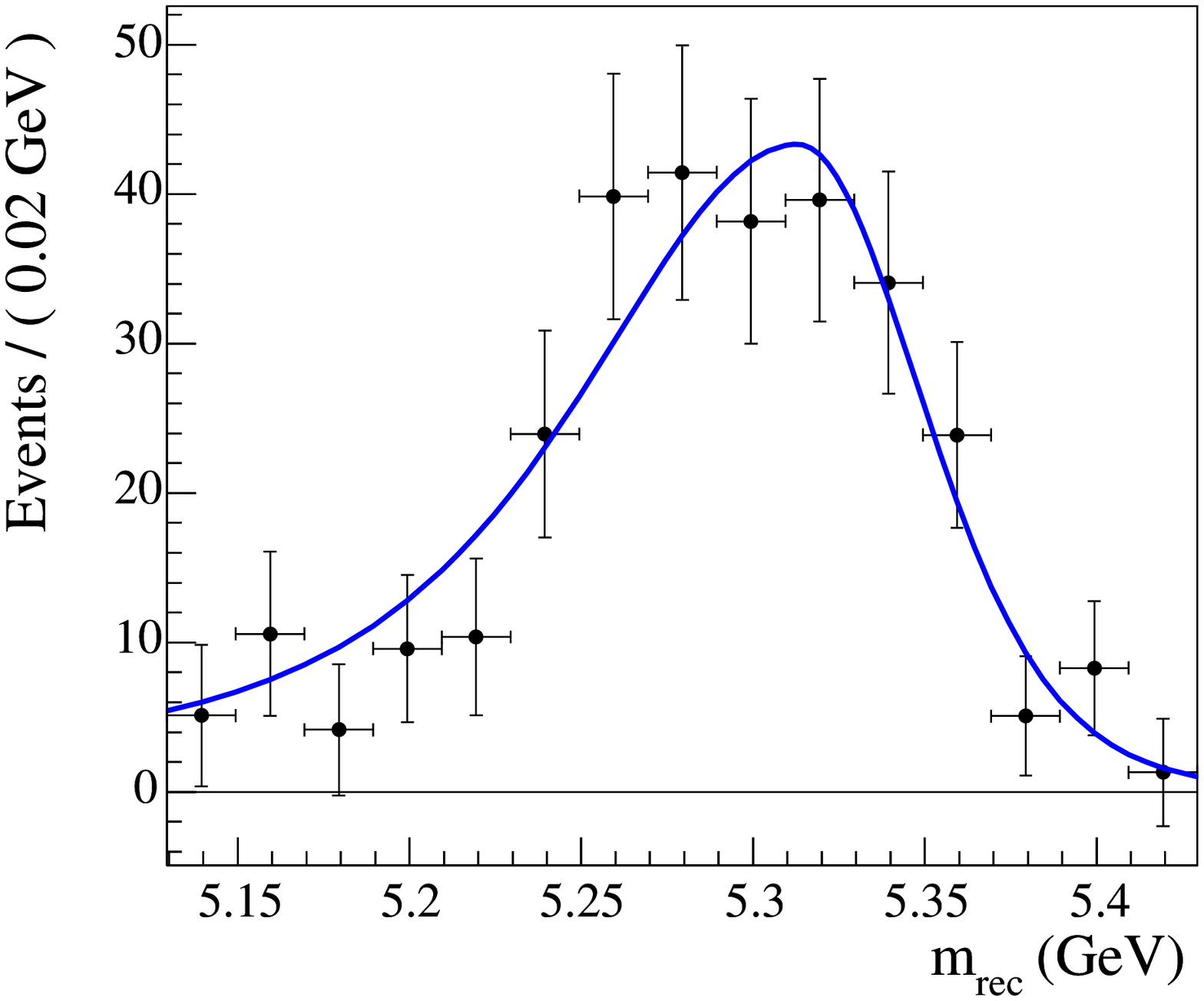}
  \includegraphics[width=0.49\linewidth]{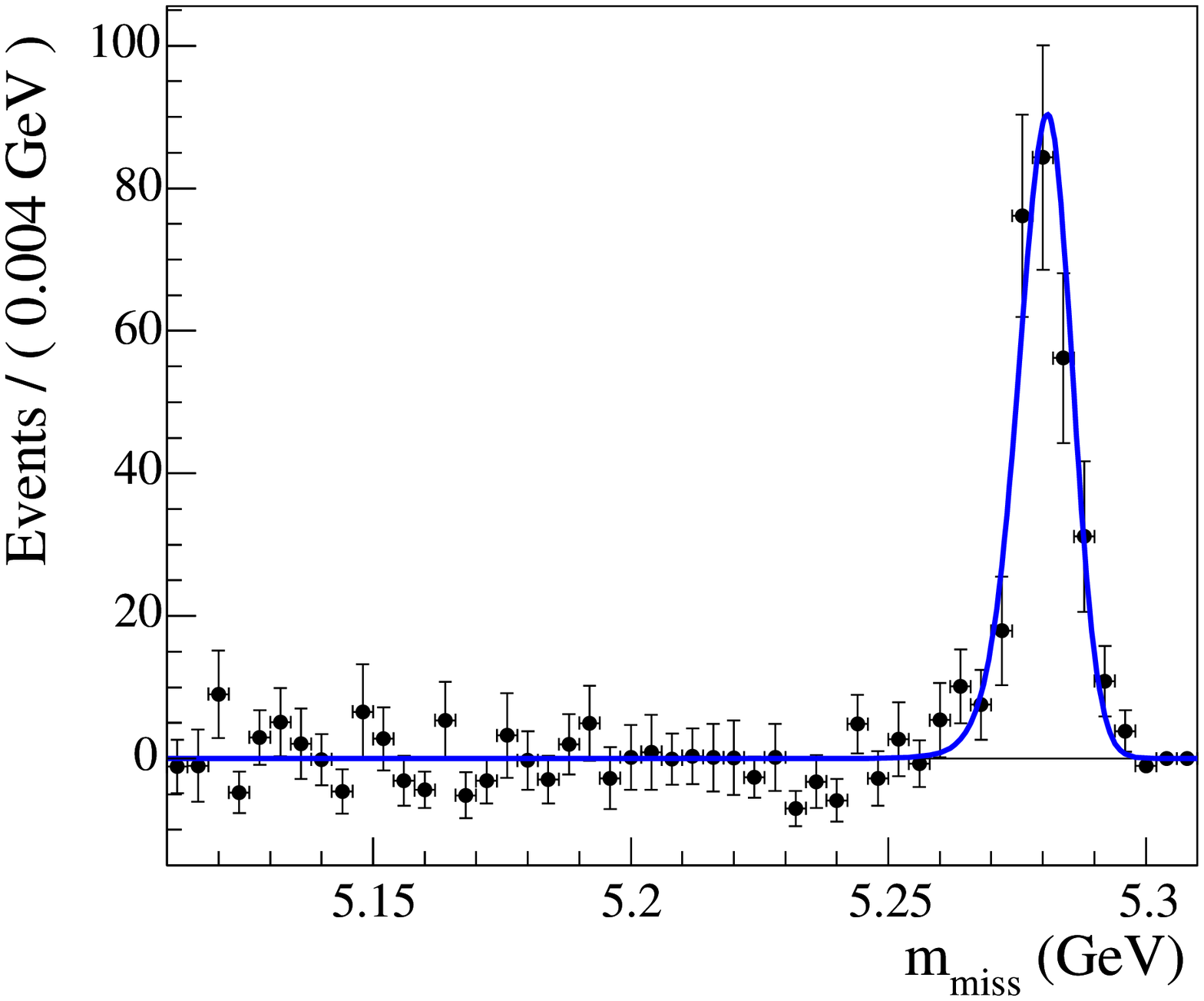}
    \begin{picture}(0,0)(0,0)
      \rput[cb](-5.5,5.9){\babarprelim}
      \rput[cb](-13.7,5.9){\babarprelim}     
    \end{picture}
  \caption{Signal \sPlot{}s for reconstructed mass (left) and missing mass (right). 
    The curve represents the PDF used for signal events in the maximum likelihood fit.}
  \label{fig:bkinsplots}
\end{figure}

\begin{figure}[!tbp]
  \includegraphics[width=0.49\linewidth]{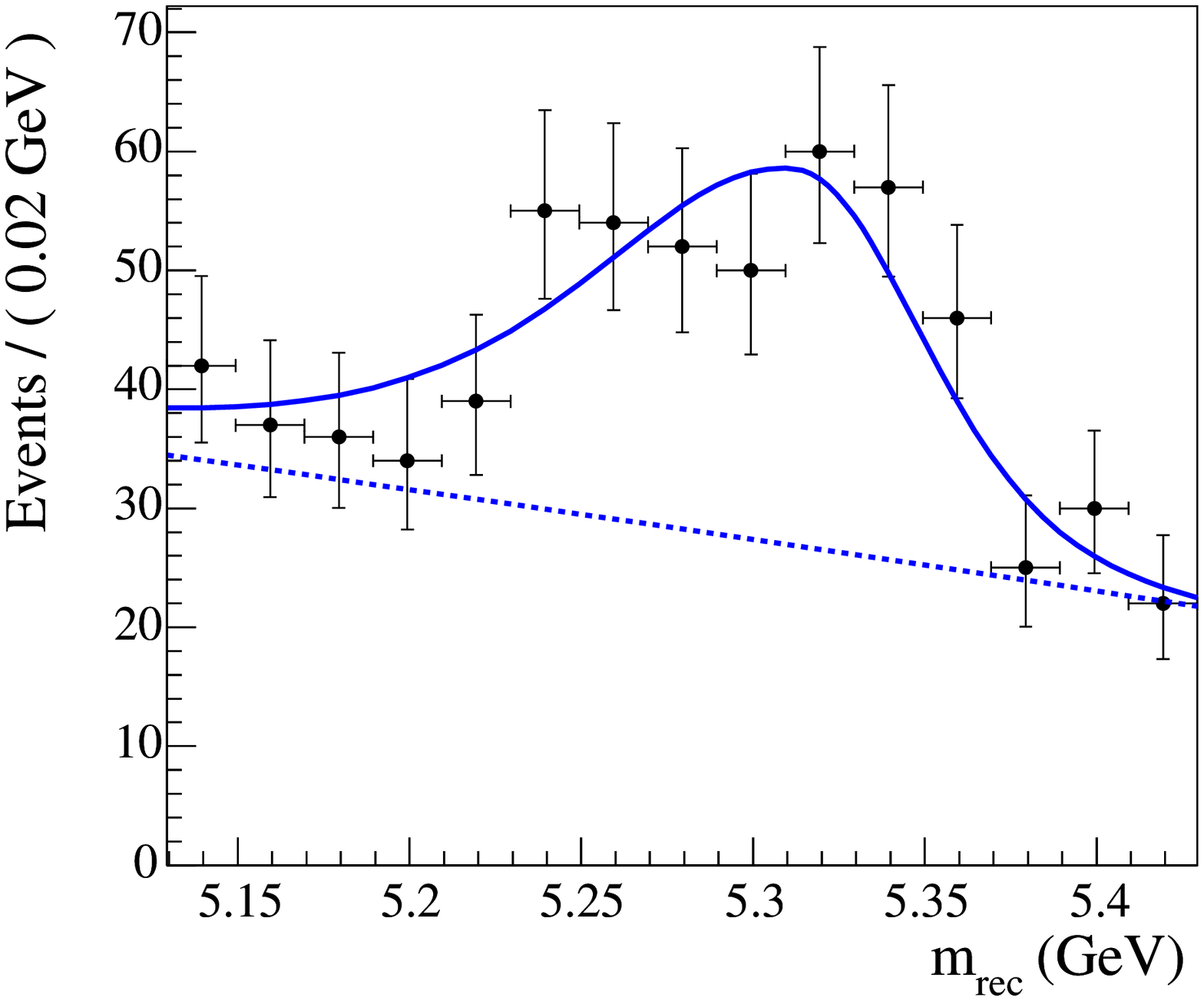}
  \includegraphics[width=0.49\linewidth]{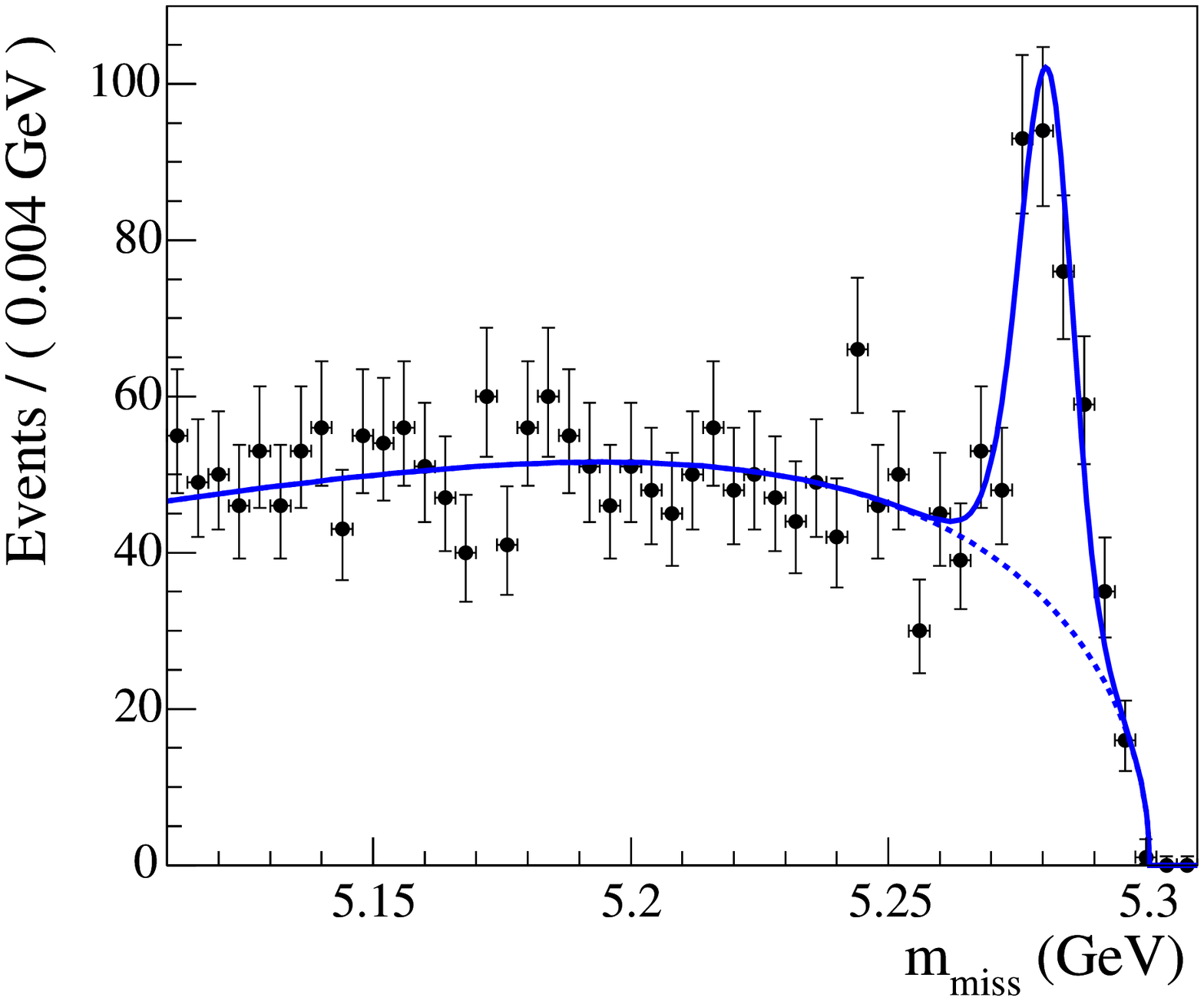}
    \begin{picture}(0,0)(0,0)
      \rput[cb](-5.5,5.9){\babarprelim}
      \rput[cb](-13.7,5.9){\babarprelim}
    \end{picture}
  \caption{Distributions for the reconstructed mass of the $B$ candidate (left)
    and for the missing mass (right). To enhance the sample in signal
    purity we required $L_{2}/L_{0}<0.4$ which reduces the efficiency
    for signal decays by \unit[$\sim25$]{\%} with respect to the
    selection described in the text. For the \mb{} distribution we
    required in addition \unit[$\mmiss>5.25$]{\gevcc}. The dashed and
    solid curves represent the background and signal-plus-background
    contributions, respectively, as obtained from the maximum
    likelihood fit.}
  \label{fig:bkinprojplots}
\end{figure}

Figure~\ref{fig:bkinsplots} shows the so called \sPlot{}
distributions~\cite{Pivk:2004ty} of the reconstructed mass and missing
mass for signal \Bztokspiz{} candidates. The \sPlot{}s are a statistical
tool to extract the distribution of a particular discriminating
variable by weighting events with their signal likelihood based on
other discriminating variables. For comparison,
figure~\ref{fig:bkinprojplots} shows the unweighted distributions for
a subsample enhanced in signal purity by selecting on $L_{2}/L_{0}$ and
\mmiss{}. The dashed and solid curves indicate background and
signal-plus-background contributions, respectively, as obtained from
the fit, but corrected for the selection.

\begin{figure}[!tbp]
  \begin{center}
    \includegraphics[width=0.9\linewidth]{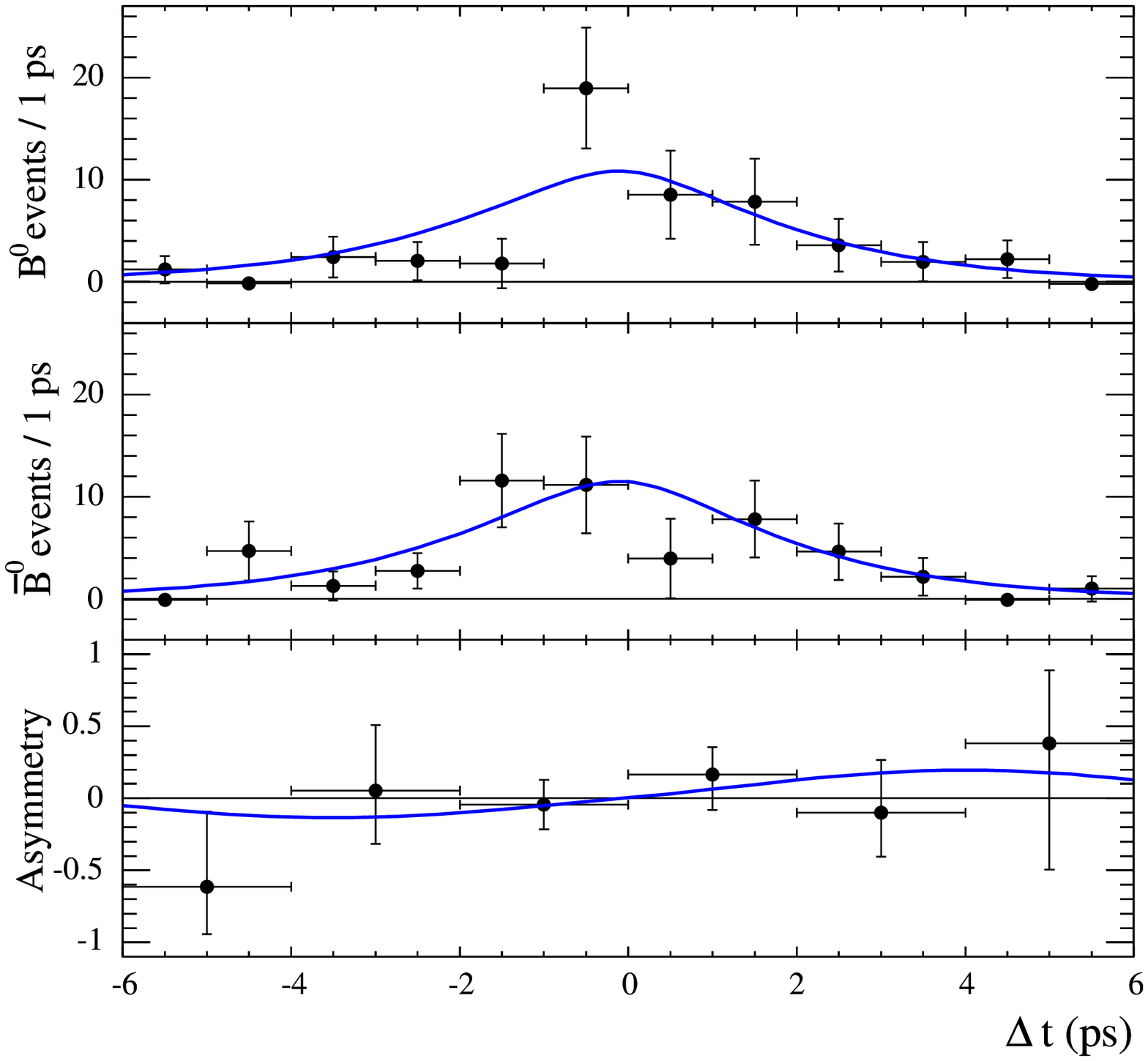}
    \begin{picture}(0,0)(0,0)
      \rput[cb](-3.,12.3){\babarprelim}
    \end{picture}
  \end{center}
\caption{\sPlot{} distributions of $\deltat$ for signal events with
  $B_{\rm tag}$ tagged as $\Bz$ (top) or $\Bzb$ (center), and of the
  asymmetry ${\cal A}_{\KS\piz}(\deltat)$ (bottom). The curves represent the
  PDFs for signal decays in the likelihood fit.}
\label{fig:dtplot}
\end{figure}

\begin{figure}[!tbp]
  \centerline{
    \includegraphics[width=0.8\linewidth]{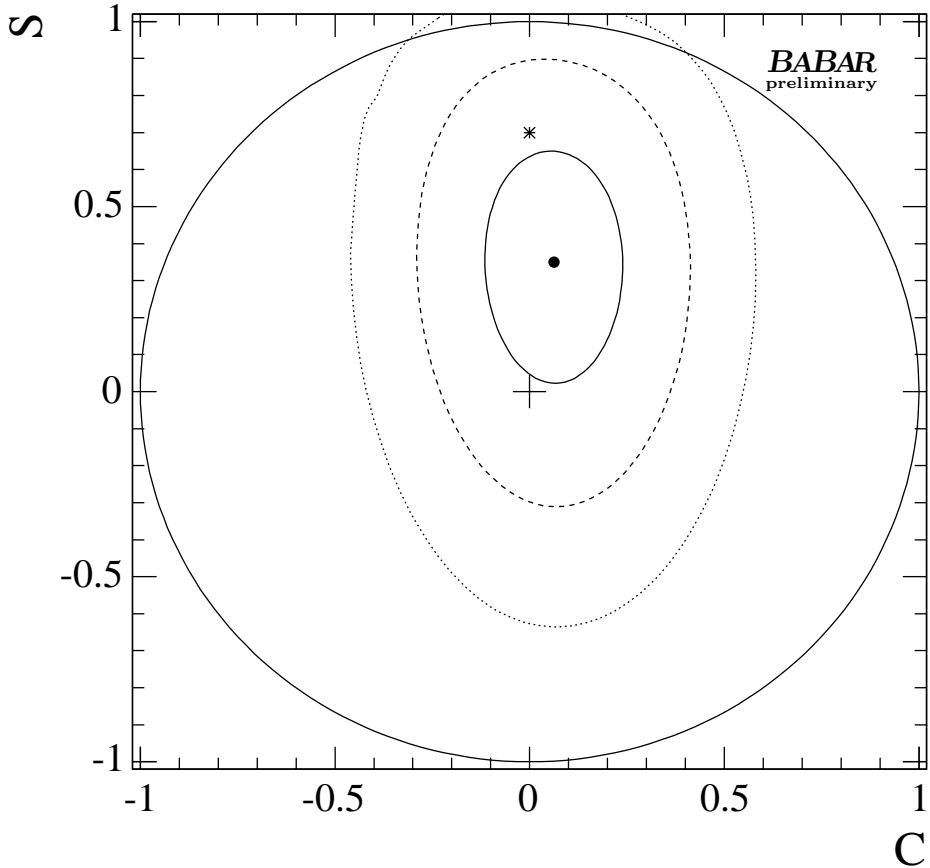}
    \begin{picture}(0,0)(0,0)
      \rput[cb](-2.2,11.3){\babarprelim}
    \end{picture}
    }
  \caption{Contours for constant likelihood corresponding to a change in the likelihood 
    with $1$, $2$ and $3$ units with respect to the minimum
    likelihood. The enclosed regions correspond roughly to
    \unit[$39$]{\%}, \unit[$86$]{\%} and \unit[$99$]{\%} confidence
    levels. The star represents the Standard Model prediction. The
    large circle represents the boundary of the physically allowed
    region.}
  \label{fig:SvsC_contour_plot_BS}
\end{figure}

Figure~\ref{fig:dtplot} shows the \sPlot{} distributions of $\deltat$
for $\Bz$- and $\Bzb$-tagged events, and of the asymmetry ${\cal
  A}_{\KS\piz}(\deltat) = \left[N_{\Bz} -
  N_{\Bzb}\right]/\left[N_{\Bz} + N_{\Bzb}\right]$ as a function of
$\deltat$. Figure~\ref{fig:SvsC_contour_plot_BS} shows the contours
for constant likelihood in the $S-C$ plane.

\section{VALIDATIONS AND SYSTEMATIC UNCERTAINTIES}

The extraction of \deltat{} with the IP-constrained geometric decay
tree fit has been extensively tested on large samples of simulated
\Bztokspiz{} decays with different values of $C$ and $S$. To evaluate
uncertainties due to the use of a resolution function extracted from
the $B_{\rm flav}$ sample we fit the \Bztokspiz{} samples with a
resolution function extracted from simulated \Bztojpsiks{} events. We
assign a systematic uncertainty of $0.02$ on \skspiz{} and $0.01$ on
\ckspiz{} due to the use of the $B_{\rm flav}$ resolution function and
other effects related to the \deltat{} reconstruction. 
We evaluate the effect of a possible misalignment of the SVT by introducing 
misalignments in the simulation and obtain a 
systematic uncertainty of $0.03$ on \skspiz{} and $0.01$ on \ckspiz.
We also consider large
variations of the position and size of the interaction region, which
we find to have negligible impact. We include a systematic uncertainty
of $0.02$ on both \skspiz{} and \ckspiz{} to account for imperfect
knowledge of the PDFs used in the fit. We assign a systematic
uncertainty of $0.05$ to \ckspiz{} due to possible asymmetries in the
rate of \Bz{} versus \Bzb{} tags in background events.

In order to exclude any bias in the IP constraint \deltat{}
reconstruction that would be specific to data only, we examine a
sample of approximately $1900$ \Bztojpsiks{} decays with
$\jpsi\to\mup\mun$ and $\jpsi\to\epem$.  In these events we determine
\deltat{} in two ways: by fully reconstructing the $\Bz$ decay vertex
using the trajectories of charged daughters of the $\jpsi$ and the
$\KS$ mesons, or by neglecting the $\jpsi$ contribution to the decay
vertex and using the IP constraint and the \KS{} trajectory only. This
study shows that within statistical uncertainties the IP-constrained
\deltat{} measurement is unbiased with respect to the more established
technique, and that the values of $S_{\jpsi\KS}$ and
$C_{\jpsi\KS}$ so obtained are consistent. In addition, we examine the pull of the
\deltat{} difference, assuming that the \deltat{} uncertainties are
fully correlated. We find that the pull distribution in the data is
approximately \unit[$10$]{\%} wider than that in the simulation. We therefore
include an additional systematic uncertainty of $0.014$ in
\skspiz{}. Finally, we measure the $\Bz$ lifetime in \Bztokspiz{}
decays and in IP-constrained \Bztojpsiks{} decays and find that both
agree with the world average.

The detection efficiency for signal events is calculated with a Monte
Carlo simulation based on the Pythia event
generator~\cite{Sjostrand:1993yb} and GEANT4 detector
simulation~\cite{Agostinelli:2002hh}. The efficiency of the \KS{}
selection is calibrated with a large sample of inclusive
$\KS\to\pip\pim$ decays. The $\piz\to\gamma\gamma$ efficiency
is calibrated with $\epem\to\taup\taum$ events with
$\tau^-\to\rhom\nu_{\tau}$. The systematic uncertainty associated with
the efficiency is \unit[$2.6$]{\%} for \KS{} and \unit[$3.0$]{\%} for
\piz{}. We assign additional systematic uncertainties of
\unit[$1.2$]{\%} for the $L_2/L_0$ cut, \unit[$2.0$]{\%} for the
selection on \mb{} and a total of \unit[$2.0$]{\%} for uncertainties in
the signal PDFs. Finally, we include an uncertainty of
\unit[$1.4$]{\%} to account for unknown contributions from other \BB{}
decays and an uncertainty of \unit[$0.6$]{\%} in the total number of
$\Y4S\to\BB$ decays.

\section{SUMMARY}

In this paper we have reported preliminary results from a measurement
of the branching fraction and time-dependent CPV asymmetries of
\Bztokspiz{} decays. The measured values of \skspiz{} and \ckspiz{}
are consistent with the Standard Model predictions. The measured
branching fraction is consistent with measurements from other
experiments~\cite{Brkspiz}.

\section{ACKNOWLEDGMENTS}

\par
We are grateful for the 
extraordinary contributions of our \pep2\ colleagues in
achieving the excellent luminosity and machine conditions
that have made this work possible.
The success of this project also relies critically on the 
expertise and dedication of the computing organizations that 
support \babar.
The collaborating institutions wish to thank 
SLAC for its support and the kind hospitality extended to them. 
This work is supported by the
US Department of Energy
and National Science Foundation, the
Natural Sciences and Engineering Research Council (Canada),
Institute of High Energy Physics (China), the
Commissariat \`a l'Energie Atomique and
Institut National de Physique Nucl\'eaire et de Physique des Particules
(France), the
Bundesministerium f\"ur Bildung und Forschung and
Deutsche Forschungsgemeinschaft
(Germany), the
Istituto Nazionale di Fisica Nucleare (Italy),
the Foundation for Fundamental Research on Matter (The Netherlands),
the Research Council of Norway, the
Ministry of Science and Technology of the Russian Federation, and the
Particle Physics and Astronomy Research Council (United Kingdom). 
Individuals have received support from 
CONACyT (Mexico),
the A. P. Sloan Foundation, 
the Research Corporation,
and the Alexander von Humboldt Foundation.


\begin{thebibliography}{99}

\bibitem{CKM}
  N.~Cabibbo, \jprl {\bf 10}, 531 (1963);
  M.~Kobayashi and T.~Maskawa, \progtp{\bf 49}, 652 (1973).

\bibitem{BaBarSin2betaObs}
 B. Aubert {\em et al.} [\babar{} Collaboration], \jprl{\bf 87}, 091801 (2001).

\bibitem{BelleSin2betaObs}
  K. Abe {\em et al.} [Belle Collaboration] , \jprl{\bf 87}, 091802 (2001).

\bibitem{Eidelman:pdg2004}
S. Eidelman {\it et al.} [Particle Data Group Collaboration], \plb{\bf 592}, 1 (2004).

\bibitem{ref:newphysics}
  Y.~Grossman and M.~P.~Worah, \plb{\bf 395}, 241 (1997).
  M.~Ciuchini, E.~Franco, G.~Martinelli, A.~Masiero and L.~Silvestrini,
  \jprl{\bf 79}, 978 (1997).
  D.~London and A.~Soni, \plb{\bf 407}, 61 (1997).

\bibitem{ref:cc} Unless explicitly stated, conjugate decay modes are
assumed throughout this paper.




\bibitem{Abe:2003yt}
K.~Abe {\it et al.}  [Belle Collaboration],
\jprl{\bf 91}, 261602 (2003).

\bibitem{Aubert:2004ii}
B.~Aubert {\it et al.}  [\babar{} Collaboration],
arXiv:hep-ex/0403026.

\bibitem{Aubert:2003bq}
B.~Aubert {\it et al.}  [\babar{} Collaboration],
\jprl{\bf 91}, 161801 (2003).

\bibitem{Aubert:2004ta} B.~Aubert {\it et al.}  [\babar{} Collaboration], arXiv:hep-ex/0406005.

\bibitem{Aubert:2004am}
B.~Aubert {\it et al.}  [\babar{} Collaboration],
arXiv:hep-ex/0406040.

\bibitem{Aubert:2004xf}
B.~Aubert {\it et al.}  [\babar{} Collaboration],
arXiv:hep-ex/0403001, submitted to \jprlBase{}

\bibitem{Fleischer:1995cg}
R.~Fleischer,
\plb{\bf 365}, 399 (1996).

\bibitem{Gronau:2003kx}
M.~Gronau, Y.~Grossman and J.~L.~Rosner,
\plb{\bf 579}, 331 (2004).

\bibitem{ref-qcdbounds}
A.~J.~Buras, R.~Fleischer, S.~Recksiegel and F.~Schwab,
arXiv:hep-ph/0402112.
M.~Ciuchini, E.~Franco, G.~Martinelli, A.~Masiero, M.~Pierini and L.~Silvestrini,
arXiv:hep-ph/0407073.
J.~Charles {\it et al.}  [CKMfitter Group Collaboration],
arXiv:hep-ph/0406184.

\bibitem{Aubert:2003sg}
B.~Aubert {\it et al.}  [\babar{} Collaboration],
\jprl{\bf 92}, 201802 (2004).

\bibitem{ref-kpipuzzle1} 
M.~Ciuchini, E.~Franco, G.~Martinelli, M.~Pierini and L.~Silvestrini,
Phys.\ Lett.\ B {\bf 515} (2001) 33.
Y.~Y.~Keum, H.~N.~Li and A.~I.~Sanda,
Phys.\ Rev.\ D {\bf 63}, 054008 (2001).
M.~Beneke and M.~Neubert,
Nucl.\ Phys.\ B {\bf 675} (2003) 333.

\bibitem{Buras:2003dj}
A.~J.~Buras, R.~Fleischer, S.~Recksiegel and F.~Schwab,
Phys.\ Rev.\ Lett.\  {\bf 92}, 101804 (2004).

\bibitem{ref:babar}
B. Aubert {\em et al.} [\babar{} Collaboration] , \nima{479}, 1 (2002).

\bibitem{ref:Sin2betaPRD}
B. Aubert {\em et al.} [\babar{} Collaboration] ,
\jprd{\bf 66}, 032003 (2002).

\bibitem{ref:sin2betaPRL02}
B.~Aubert {\it et al.} [\babar{} Collaboration] ,
\jprl{\bf 89}, 201802 (2002).


\bibitem{Pivk:2004ty}
M.~Pivk and F.~R.~Le Diberder,
arXiv:physics/0402083.

\bibitem{Sjostrand:1993yb}
T.~Sjostrand,
Comput.\ Phys.\ Commun.\  {\bf 82}, 74 (1994).

\bibitem{Agostinelli:2002hh}
S.~Agostinelli {\it et al.}  [GEANT4 Collaboration],
Nucl.\ Instrum.\ Meth.\ A {\bf 506}, 250 (2003).

\bibitem{Brkspiz}
  B.~C.~K.~Casey {\it et al.}  [Belle Collaboration],
  Phys.\ Rev.\ D {\bf 66}, 092002 (2002).
  A.~Bornheim {\it et al.}  [CLEO Collaboration],
  Phys.\ Rev.\ D {\bf 68}, 052002 (2003).




\end{thebibliography}
\end{document}